# Path Loss Model Modification for Various Gains and Directions of Antennas

Jan M. Kelner and Cezary Ziółkowski
Institute of Telecommunications, Faculty of Electronics, Military University of Technology, Warsaw, Poland
{jan.kelner, cezary.ziolkowski}@wat.edu.pl

*Abstract*—Emerging telecommunication technologies like 5G, green communications, and massive MIMO contribute to the use of directional antennas and beamforming. For this reason, modern propagation models should consider directional antennas with different beam widths, gains and radiation pattern directions. Most of the available propagation models are based on omnidirectional or isotropic antennas. A proposition to solve this problem is an empirical path loss model that considers the different types of antennas. This model assumes that the transmitting and receiving antenna beams are directing each other. In this paper, we propose modifying this model by determining attenuation for any direction of the antenna patterns. For this aim, the multi-elliptic channel model is used.

## I. Introduction

Propagation models can be grouped by three main types, i.e., large-, medium- and small-scale models. The large-scale models represent changes in a signal attenuation or field strength versus a distance from a transmitter (Tx). The medium-scale models describe shadowing effect. The small-scale models, also called channel models, reflect other physical phenomena occurring in radio channels such as delays, scattering, fading, or Doppler effect.

Most analytical and simulation models for large and small scales assume the use of omnidirectional or isotropic antennas. The same applies to standard models, also called reference model, that based on empirical measurements in real propagation environments, such as 3GPP [1], COST, WINNER [2]. In these models, different scenarios are defined by cluster delay line (CDL) or tapped delay line (TDL) models. TDL is usually represented by a set of powers and delays corresponding to a typical power delay profile (PDP). The CDL contains additional information on the shape of the power angular spectrum (PAS), sets of angles of departure (AODs) and angles of arrival (AOAs) for each TDL time cluster.

In the era of a beamforming technology development for 5G, multi-input multi-output (MIMO) and massive MIMO systems, or green communications, the adaptation of directional antennas is becoming increasingly important. Thus, the propagation models should also be able to include antennas with different gains and beam widths. In some cases, special modifications of models are proposed. For example, a spatial TDL filtering is used in the 3GPP model [1]. However, the results obtained in this way do not accurately reflect the phenomena occurring in the real multipath propagation environment.

Empirical models developed for specific types of antennas provide a better fit to the measurement results. An example of this may be the large scale model developed by J. A. Azevedo, F. E. Santos, T. A. Sousa, and J .M. Agrela, so called the ASSA model (ASSAM) [3]. This path loss model is parameterized for a dozen antenna types. However, ASSAM only considers the spatial situation where the antenna beams are directing each other. It is a disadvantage of this solution.

The purpose of this paper is to present a numerical method for modifying the ASSAM, which considers the different directions and half-power beam widths (HPBWs) of the transmitting and receiving antennas. In this case, the authors used the multi-elliptical channel model (MCM) [4],[5]. We assumed that HPBW in the elevation plane is small. Hence, the ASSAM modification shown in this paper only takes the analysis in the azimuth plane into account. The MCM presented in [4],[5] is 3D model, so, in the near future, the authors also plan to include the elevation plane.

The paper is organized as follows. In Section II, the short description of ASSAM is presented. The MCM used to modify the path loss model is briefly described in Section III. Section IV shows the method of modifying the ASSAM. Sample results for different directions and beam widths of antennas are presented in Section V. The summary is in Section VI.

## II. Path Loss Model for Various Gains and Directivity of Antennas

ASSAM [3] is the path loss model based on empirical measurements for two environments: campus and forest. Attenuation measurements were made up to 150 m from the transmitter (Tx) for a dozen antenna types. Based on the measurements, path loss exponents for each antenna type are determined and presented in [3, Table 1]. In ASSAM, the path loss at a distance, $d$, from Tx for the analyzed antenna type is defined as

$$\mathrm{PL}_0(d)(\mathrm{dB}) = \mathrm{PL}(d_0)_{\mathrm{ref}} + 10n\log_{10}\left(\frac{d}{d_0}\right) \quad (1)$$

where $\text{PL}(d_0)_{\text{ref}}$ is the path loss measured by a reference antenna at a reference distance, $d_0$, and $n$ is the path loss exponent for the analyzed antenna type.

In [3], the half-wave dipole is adopted as the reference antenna and $d_0 = 5$ m is the reference distance. So, based on [3, Table 1], we assume $\text{PL}(d_0)_{\text{ref}} = 53.1$ dB.

ASSAM is valid for 2.4 GHz and distance range from 5 to 400 m. The model includes a dozen types of antennas with different HPBWs in azimuth and elevation planes, and gains from 3 to 46 dBi [3, Table 1]. Furthermore, based on ASSAM, the linear relationship between the received signal strength and the antenna gain is shown.

### III. MULTI-ELLIPTICAL CHANNEL MODEL CONSIDERING A INFLUENCE OF ANTENNA PATTERNS

MCM is based on the Parsons-Bajwa model [6], so Tx and receiver (Rx) are located in the ellipse foci. The ellipses define the potential location of the scatterers of signal components. MCM spatial structure in the azimuth plane is shown in Fig. 1. In relation to the Parsons-Bajwa model, apart from the delayed scattering components and the direct path component, MCM also includes the local scattering components occurred around the Rx antenna [7],[8]. For this aim, the von Mises distribution is used. Empirical or standard model-based PDPs are used to define delayed scattering components and thus the ellipse sizes.

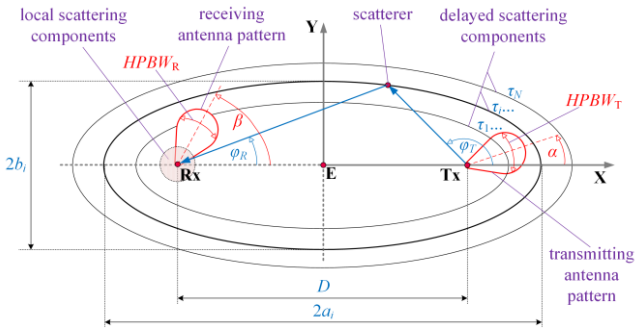

Fig. 1. Spatial structure of MCM.

MCM shown in [7],[8] only includes the omnidirectional antennas in the azimuth plane. 2D MCM including directional Tx antenna is shown in [9]. 3D MCM extensions including the elevation plane, the transmitting and receiving antenna patterns are presented in [4] and [5], respectively. In the case of the omnidirectional antenna, a uniform distribution is used to model its radiation pattern. In the case of the directional antenna, a Gaussian with the corresponding HPBW is used to model the main lobe of each of the antenna pattern.

A distribution describing the Tx antenna pattern is used to define a probability density function (PDF) of AOD for delayed scattering components [4],[9]. In addition, the Tx antenna pattern modifies the power of the first PDP time-cluster, which corresponds to the local scattering components and/or direct path component. The distribution describing the Rx antenna pattern is used to modify the powers of the received signal components [5].

In the paper, generating PAS with considering the Tx and Rx antenna patterns is implemented as in [4],[5], but only with respect to the azimuth plane. This approach is presented in [10].

### IV. ASSAM MODIFICATION

ASSAM is the large-scale model for different types of overlapping antennas directing each other, but it does not provide an ability to evaluate the attenuation for other antenna orientations. However, MCM gives an opportunity to assess the influence of the antenna patterns and their directions on PAS and the total power of the received signal. This MCM property is used in the ASSAM modification.

The path loss in free space, for cases where the antennas are directed at themselves ($\alpha = 180°, \beta = 0°$), can be defined as

$$\text{PL}_0(\text{W/W}) = \text{PL}(\alpha = 180°, \beta = 0°) = \frac{P_T G_T G_R}{P_R} \quad (2)$$

where $P_T$ and $P_R$ are the powers at the Tx output and Rx input, respectively, $G_T$ and $G_R$ are the Tx and Rx antenna gains, respectively.

In case the antennas are directed in different directions, you need to enter a correction factor, $K$, for the received signal power. Then the path loss can be described as

$$\text{PL}(\alpha, \beta) = \frac{P_T G_T G_R}{K(\alpha, \beta) P_R} = \frac{\text{PL}(\alpha = 180°, \beta = 0°)}{K(\alpha, \beta)} \quad (3)$$

Thus, for a generalized case, we obtain the equation in dB

$$\text{PL}(\alpha, \beta)(\text{dB}) = \text{PL}_0(\text{dB}) - 10 \log_{10} K(\alpha, \beta) \quad (4)$$

where $\text{PL}_0 = \text{PL}_0(d)$ is the path loss at the distance from Tx, $d$, based on ASSAM for the analyzed antenna type, $\text{PL}(\alpha, \beta) = \text{PL}(d, \alpha, \beta)$ is the modified attenuation determined at the same distance for $\alpha$ and $\beta$ directions of the Tx and Rx antennas, respectively.

The correction factor, $K(\alpha, \beta)$, of the power seen at the Rx input is determined on the basis of MCM. In this case, for a given distance Tx-Rx, two PASs are determined, $\text{PAS}_0(\varphi) = \text{PAS}(\varphi, \alpha = 180°, \beta = 0°)$ and $\text{PAS}(\varphi) = \text{PAS}(\varphi, \alpha, \beta)$ for two configurations of the analyzed antenna type, according to the methodology presented in [4],[5]. On their basis, the total received powers and correction factor are determined, as follows

$$P_0\left(\alpha=180°,\beta=0°\right)=\int_{-\pi}^{\pi}\mathrm{PAS}\left(\varphi,\alpha=180°,\beta=0°\right)\mathrm{d}\varphi \quad (5)$$

$$P(\alpha,\beta)=\int_{-\pi}^{\pi}\mathrm{PAS}(\varphi,\alpha,\beta)\mathrm{d}\varphi \quad (6)$$

$$K(\alpha,\beta)=\frac{P(\alpha,\beta)}{P_0\left(\alpha=180°,\beta=0°\right)} \quad (7)$$

This coefficient is determined for the analyzed distance and direction of the Tx and Rx antennas, and substituted into (4).

## V. PATH LOSS VS ANTENNA DIRECTIONS

The evaluation of the ASSAM modification is performed for the two antenna types shown in [3], i.e., the corner reflector (CR) and parabolic grid (PG). Based on [3, Table 1], the path loss exponents are $n_{\mathrm{CR}}=5.28$ and $n_{\mathrm{PG}}=7.08$, while the beam widths in the azimuth plane are $HPBW_{\mathrm{CR}}=58°$ and $HPBW_{\mathrm{PG}}=10°$ for CR and PG, respectively.

Using (1), the attenuation is calculated for the analyzed antenna type and the distance range from 10 to 400 m. Based on MCM, the correction factors, $K(\alpha,\beta)$, are obtained for the each analyzed distance, $\alpha$ and $\beta$ directions of the antennas. These data are substituted into (4).

PDP as an input for MCM is required. Because PDPs for ASSAM-based measurements are not available in [3], the authors used PDP from the standard model. In this case, PDP from the 3GPP model is used, which corresponds to the TDL-B [1, Table 7.7.2-2] for Urban Macro (UMa) scenario, non-line-of-sight (NLOS) conditions, carrier frequency $f_c=2.4\,\mathrm{GHz}$, and rms delay spread, $DS=363$ ns [1, Table 7.7.3-2].

Figures 2 and 3 show $K$ and PL versus $d$ for CR, $\alpha=180°$, and selected $\beta$, respectively. Analogous graphs, for $\beta=0°$ and selected $\alpha$, are shown in Figs. 4 and 5.

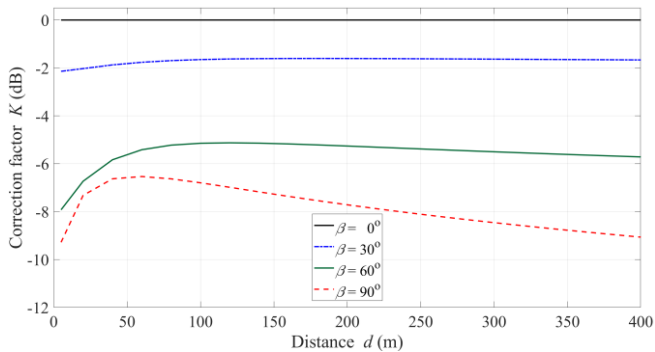

Fig. 2. Correction factor versus distance for CR, $\alpha=180°$ and selected $\beta$.

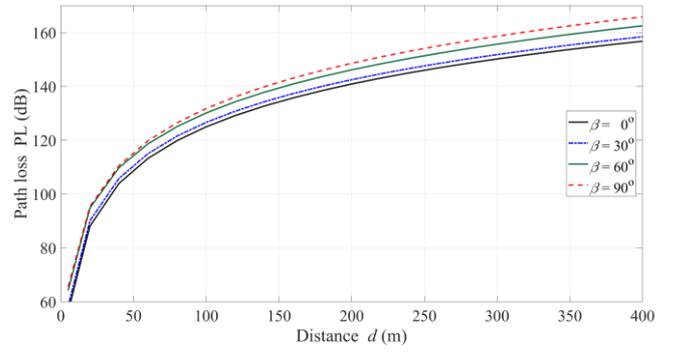

Fig. 3. Path loss versus distance for CR, $\alpha=180°$ and selected $\beta$.

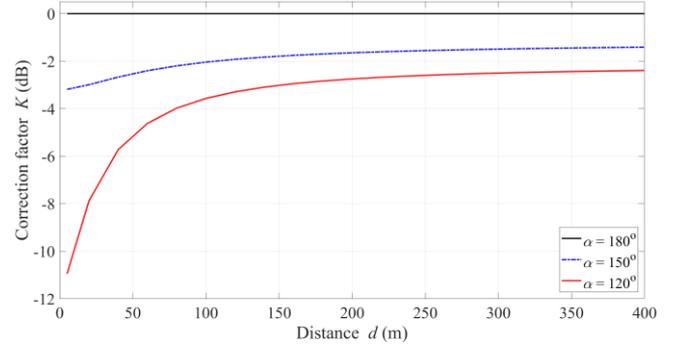

Fig. 4. Correction factor versus distance for CR, $\beta=0°$ and selected $\alpha$.

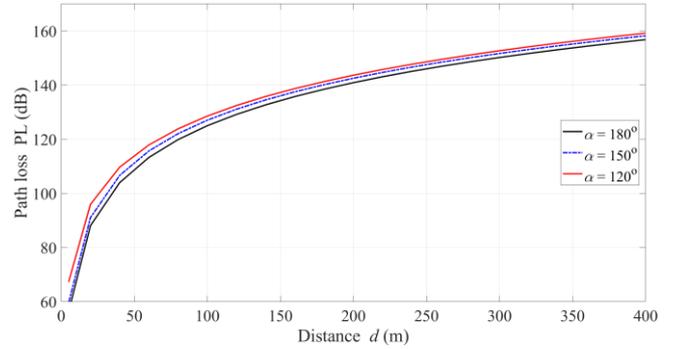

Fig. 5. Path loss versus distance for CR, $\beta=0°$ and selected $\alpha$.

The correction factor decreases with increasing $|\beta|$ for $\alpha=180°$ and with decreasing $|\alpha|$ for $\beta=0°$. This effect results in an increase of the path loss with a change in antenna directions from the orientation $\alpha=180°$ and $\beta=0°$. In general, for all sectoral or directional antenna type, the path loss changes are higher for changing the direction of the receiving antenna than the transmitting antenna. Simultaneous changes of $\alpha$ and $\beta$ may result in larger changes in the correction factor and path loss, as shown in Fig. 7.

For two types of antennas, i.e., CR and PG, the influence of the antenna beam width on the attenuation is shown in Figs. 6 and 7. Figure 6 presents the path loss for $\alpha=180°$ and selected $\beta$, while Fig. 7 shows graphs for simultaneous changes of $\alpha$ and $\beta$.

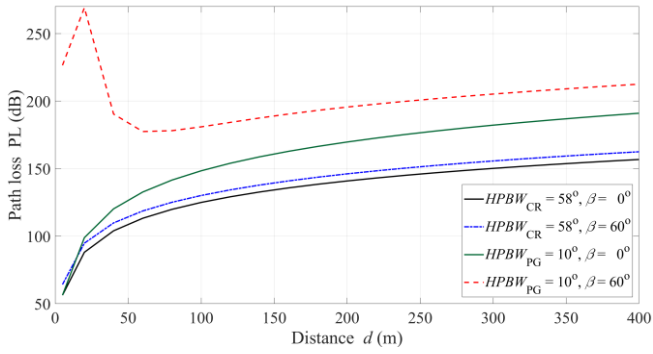

Fig. 6. Path loss versus distance for CR and PG, $\alpha = 180°$, and selected $\beta$.

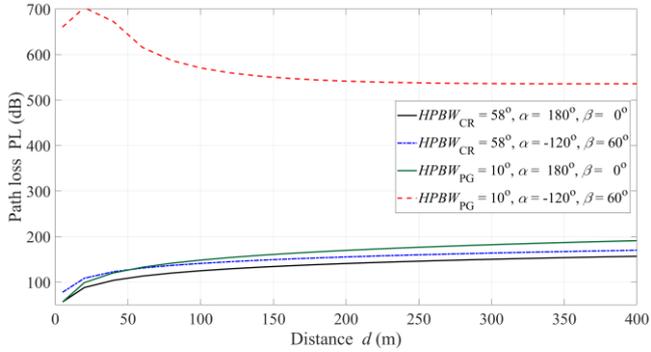

Fig. 7. Path loss versus distance for CR and PG, selected $\alpha$ and $\beta$.

Reducing HPBW provides increasing the path loss. An additional change in the orientations of the antennas causes higher attenuation for smaller than larger HPBW.

In Figs. 6 and 7, for certain orientations of the PG antennas, we first see the increase and then the decrease of the path loss along with the distance. This case occurs mainly for the narrow-beam antennas and is caused by "spatial filtration" by the receiving antenna pattern. Examples of suitable PASs are shown in [10, Figs. 6,7].

The obtained results indicate the importance of the proper directions of the transmitting and receiving antennas, especially in the cases of the selective directional antennas. It is really important for microwave relay links in backhaul, beamforming for 5G and green communications.

## VI. CONCLUSION

In the paper, the methodology of modifying the path loss model for different antenna directions has been shown. This modification is based on MCM and was used for ASSAM. ASSAM is the empirical large-scale model that consider antennas directed at themselves. Use of methodology allows to determine the attenuation for other directions of the transmitting and receiving antennas. Obtained results indicate significant influence of the antenna orientations on the received signal attenuation. The presented modification method can also be applied to other models that include, e.g., isotropic or omnidirectional antennas.